\documentclass[runningheads]{llncs}

\usepackage[T1]{fontenc}
\usepackage{graphicx}
\usepackage{mdframed}
\usepackage{todonotes}

\begin{document}

\titlerunning{Proximity-Based Indexing}
\title{Known by the Company it Keeps: \\ Proximity-Based Indexing for \\ 
Physical Content in Archival Repositories}

\author{Douglas W. Oard\orcidID{0000-0002-1696-0407}}
\institute{University of Maryland, College Park, USA \\
\email{oard@umd.edu}}

\authorrunning{D. Oard}

\maketitle

\begin{abstract}
Despite the plethora of born-digital content, vast troves of important content remain accessible only on physical media such as paper or microfilm.  The traditional approach to indexing undigitized content is using manually created metadata that describes it at some level of aggregation (e.g., folder, box, or collection).  Searchers led in this way to some subset of the content often must then manually examine substantial quantities of physical media to find what they are looking for.  This paper proposes a complementary approach, in which selective digitization of a small portion of the content is used as a basis for proximity-based indexing as a way of bringing the user closer to the specific content for which they are looking.  Experiments with 35 boxes of partially digitized US State Department records indicate that box-level indexes built in this way can provide a useful basis for search.
\keywords{Proximity-based indexing \and Archival access \and Physical media}
\end{abstract}

\section{Introduction}

The storyteller Aesop wrote (in Greek) what has been translated as “a man is known by the company he keeps”~\cite{aesop}. In sociology, Aesop’s claim is reflected in the concept of homophily, a dictionary definition of which is ``the tendency to form strong social connections with people who share one’s defining characteristics, as age, gender, ethnicity, socioeconomic status, personal beliefs, etc.''~\cite{mcpherson2001birds}.  Homophily arises in many contexts, including between people in social network analysis, in organizational dynamics~\cite{ertug2022does}, and more metaphorically in, for example, the meaning of terms in natural language processing~\cite{brunila2022company}.   

Our thesis in this paper is that a form of homophily is to be expected among the content found in archival repositories.  Our basis for this is that archivists respect the original order of archival content when performing arrangement and description~\cite{prom2013archival}.  In the arrangement task, archivists organize physical materials, typically by placing those materials in folders, placing those folders in boxes, and grouping those boxes into series.  Archival materials are generally the records of some organization or individual.  Respecting the original order of those records can help to preserve the evidence of the creator’s activities that is implicit in that original order.  Because archivists consider the evidentiary value of records to be on par with their informational value, this is an important consideration.  A second consideration that argues for respecting original order is that doing so makes it possible to open collections for research use with a minimum of work on the archivist’s part.  Because that original order was useful to the organization or individual that created the records, it is reasonable to assume that users of an archive who put in the effort to learn how a particular collection is ordered will find some value in that ordering~\cite{wiedeman2019historical}.

It is this respect for original order in archival arrangement that produces the homophily that we leverage in this paper.  Specifically, we hypothesize that if we know something about the content of some records in some archival unit (e.g., folder, box, series, or repository) then we can make some plausible inferences about where certain other records that we have not yet seen might be found.  However, it is one thing to reason from first principles that such a claim might be true, and quite another thing to show that such a claim actually is true.  In this paper, we show that the claim is true in one specific case, and thus that it could be broadly true, although we leave investigation of the broader question of how widely applicable our claim is for future work.

\section{Related Work}

Rapid growth in digital content over the last half century has resulted in the initial trickle of digital content reaching archival repositories now becoming a flood.  A broad range of tools can be used to find born-digital content, and there has been considerable innovation in that space (e.g.,~\cite{liusals2015sphere,o2019future}).  Many such tools could also be used to find content digitized from physical media, such as paper or microfilm, but problems of cost and scale limit the scope of digitization efforts.  For example, in the first five months of 2023 the National Archives and Records Administration (NARA) digitized 13 million pages from their holdings of 11.7 billion pages~\cite{nara}.  Even at that impressive rate, 121,000 pages per day, it would take 375 years to digitize the paper holdings of that one repository.  Clearly, the problem of finding things on paper will not be going away any time soon.

The first problem faced by someone wishing to find materials on paper is knowing where to look.  Citations in the scholarly literature play a particularly prominent role in this process.  For example, Tibbo found that 98\% of historians followed leads in the published literature~\cite{Tibbo03a}, and Marsh, et al. found that for anthropologists 73\% did so~\cite{marsh2023attitudes}.  There are also tools that support search of descriptions created by archivists across multiple repositories, such as ArchiveGrid~\cite{falk2017archivegrid}. Once a user knows where to look, their next challenge is to learn how to find what they want there.  As Tibbo notes, it is common for scholars to write to or call archivists before visiting a repository~\cite{Tibbo03a}. Scholars also make use of finding aids that have been created by archivists to describe (among other things) the nature of the content in a collection, and how that content is arranged and described. Although full-text search of finding aids, which for example ArchiveGrid provides, can be useful, in recent years the use of a metadata format called ``Describing Archives: a Content Standard'' (DACS) has emerged as an alternative basis for searching the results of the descriptions that archivists create~\cite{dacs}.

One limitation of these approaches for finding content on physical media is that they depend entirely on descriptions that are created by archivists.  However, the same cost and scale pressures that limit digitization also limit the creation of detailed descriptions~\cite{greene2005more,trace2022archival}. Marsh, for example, notes that of 314 collections in the Smithsonian Institution's National Anthropological Archives, only 25\% had an online finding aid as of 2019~\cite{marsh2019driven}.  A second limitation is that, as Cox has pointed out, these methods for helping people find archival content were originally designed with scholars in mind, but the general public also makes extensive use of resources found in archives (e.g., for genealogical research), and such users might well need different types of support~\cite{cox2007machines}.   For both of these reasons, we see value in creating techniques to guess where specific materials that have not yet been digitized (or otherwise richly described at the level of individual items) might be found.  That is the focus of our work in this paper.

\section{The ``Subject-Numeric Files''}

In the United States, the Department of State is responsible for management of foreign relations.  Between 1963 and 1973, State maintained its records on paper as “Subject-Numeric Files”~\cite{subjectnumeric}.  Simplifying somewhat, in this filing system the top-level category is one of 56 three-letter ``primary subject'' codes (e.g., POL for Political Affairs \& Relations), the second-level category indicates a Country (e.g., Brazil), and the third-level category is a numeric code, the meaning of which is specific to each primary subject (e.g., for POL, numeric code 15-1 designates the executive branch of government, and 27-12 designates war crimes). The entire collection includes about 8.6 million pages, held by the United States National Archives and Records Administration (NARA) in College Park, Maryland.

In recent years, Brown University engaged in large-scale digitization of records that shed light on Brazilian politics.  As one part of that, Brown arranged for about 14,000 items in NARA’s Department of State Subject-Numeric Files to be digitized, all from the POL-Brazil section of those records.  They represent parts of the content of a total of 52 boxes.
NARA intends to make these records available online, although the links from the NARA catalog to most of these records are not presently working.  Fortunately, the Brown University Library makes almost all of the digitized content from 36 of those boxes available,\footnote{https://library.brown.edu/create/openingthearchives/en/} importantly using the same box identifiers.  We wrote a crawler to download up to 100 of the records from each of 35 of those boxes (the 36th box had only two digitized files, too few to be useful for our experiments).\footnote{The documents that we downloaded were the first (at most) 100 that Brown showed on the results page for each box; results pages were ordered alphabetically by Brown University's title metadata. 100 was more than enough for our experiments, so there was no need to increase the server load by crawling more documents.}  We also crawled Brown University's title metadata for each downloaded document.\footnote{Brown University's title metadata is often more concise than NARA’s title metadata for the same document, which for example sometimes also indicates document type.}  About 2\% of the PDF files that we downloaded were not actually documents but rather forms that indicated that a document was not available for scanning; we manually removed all such cases that could be identified (either by the word “Withdrawal” in the title metadata, or by viewing PDF files that were small enough---less than 400kB---to suggest that they might be a single page).  

This process resulted in 3,205 PDF documents, organized by their original location at NARA in one of 35 boxes.  The smallest number of documents per box was 22 (box 1925); the largest number was 100 (for eleven of the boxes).  The 35 box numbers are grouped in 8 numeric sequences (1900--1908, 1925--1934, 1936--1938, 1941--1944, 2129, 2131--2132, 3832--3835, 3837--3838). 
Boxes in the NARA's Department of State Subject-Numeric Files have no identifying metadata beyond the box number, but a box consists of (typically 3 to 6) folders that hold the actual documents. Brown University metadata includes the label for the folder in which a document was found, so we crawled that metadata as well.  We can therefore describe a box by the union of its folder labels.  For example, box 1902 contains folders with the following labels: 

\begin{verbatim}
POL 2-3 BRAZ 01/01/1967 
POL 5 BRAZ 01/01/1967
POL 6 BRAZ 01/01/1967
\end{verbatim}

\section{The ``BoxFinder'' Experiments}

The PDF files created by Brown University are searchable, which means that finding a \underline{digitized} document can be done with any full-text search system, and Brown University provides such a service.  The situation is quite different, however, for content from those same boxes that has not yet been digitized.  When finding \underline{undigitized} content is the goal, as is our focus in this paper, all that a user of NARA’s archive would have is folder labels.  They would need to request every box containing any folder labeled with with a subject-numeric code and date related to their search goal.  This is a slow process, since it takes NARA several hours to deliver a requested box to a user of the archives in the reading room, and it can easily take hours to examine the records in just one box.

Our ultimate goal is to accelerate this process by recommending to a user of the archive what box they should look in.  We imagine they might use what we will build one of two ways.  In the first, they use it like Google---they type in a query, and we recommend a box.  In the second, they are looking at some documents, and we recommend a box that we expect contains similar documents. 

Whichever type of query we get, we built the box index in the same way.  We pick a few digitized documents from each box, then use the OCR text from those documents to create an index that can be used to search for a box.  The way we do this is straightforward – we take all the OCR words from some number of pages, starting at the front (e.g., just the first page, or the first two pages) from some number of PDF documents (e.g., 3 documents) that we know actually are in each box.  This gives us an index in which there are 35 items that can be found (the 35 boxes), each of which is represented by a single long string of words.  If we can use this index to find other documents that are in that same box, then we will have shown that homophily is a useful basis for search, and that a document in this collection can to some extent be ``known by the company it keeps.''

\begin{figure}[t]
\begin{small}
\begin{mdframed}
\begin{verbatim}
Attached note from the Brazilian Embassy
Background checks
Biographic Information: Vistor Jose Faccione, State Deputy (ARENA) in 
    Rio Grande do Sul
Biographic Reporting
Brazil Opts to Counter Poor Image Abroad
Brazilian Laws
Church Newspaper Calls for Democracy
Closure of Brazilian Radio Stations
Daily Media Reaction Report to ITT-Chile Situation
Detention of Fernand Legros
Disability and Death of a President
Ernesto Geisel in Rio
Ester Ferraz
Foreign Minister Gibson Speaks on Brazil's Foreign Policy
GOB Announces Development Goals and Programs for 1970-73
Ineligibilities Again
Itamaraty's Role in Formulation of Brazilian Foreign Policy
Law on Party Loyalty Expected Soon
Monthly Trends Report - December 1971
\end{verbatim}
\end{mdframed}
\end{small}
\caption{Some examples of short queries built from title metadata.}
\label{fig:shortqueries}
\end{figure}

\subsection{Query Formulation}

In our experiments, we don’t have a real user, so we simulate the two search scenarios.  To simulate a ``type a query'' scenario, we search using Brown University's title metadata for some document as the query (one we have an image of, but that we had not chosen to index).\footnote{Brown University's title metadata is human-generated; many documents do not actually have titles within the document, and for those that do Brown's title metadata sometimes contains contextual terms missing from the document's actual title.}  The resulting queries have an average length of 6 terms (min 1, max 26); Figure~\ref{fig:shortqueries} shows some examples.  If our system can guess which box contains the document from which we got the title, then we expect that it could also do well if a real searcher ever typed a query like that.  
Of course, searchers might type queries that are better or worse than the document title that we used, but at least this will indicate whether our homophily-based approach can work when it gets a query like the one we gave it.  

\begin{figure}[t]
\begin{small}
\begin{mdframed}
\begin{verbatim}
Department o 
PAGE 01 
47 
ACTION SS 7* 
INFO CIAE 00,/070 W RIO DE 01693 06I913Z LIMITED OFFICIAL USE 
I I 9070 
R 06I6T0Z MAR 69 
FM AMEMPASSY RIO HE JANFIRO 
TO SECSTATE WASHDC 7226 
LIMITED OFFICIAL USE RIO DE JANEIRO t693 
L I MO IS 
FOR VAKY 
SUBJECTi DELIVERY OF ROCKEFELLER LETTER 
I, IN RFSPONSE TO REQUEST FOR AUDIENCE WITH PRES WE WERE INFORMED 
THAT IT NOT CUSTOMARY HERE FOR PRES TO PECEIVF CHARGE D'AFFAIRES 
AND PERHAPS I COULD DELIVER IT VIA FONMIN. IT WAS STATED* HOWEVER. 
THAT EXCEPTION WOULD BE MADE IN CASE OF U.S. CHARGE IF I So 
REQUESTED. 
p. IN VIEW FACT I HAD NO OTHER MAJOR SUBSTANTIVE BUSINESS TO DIS-
CUSS AND NOTHING T0 ADD TO CONTENTS OF LETTER, I ELFCTED Tn USE 
FONMIN CHANNEL FOR DELIVERY* ON THEORY THERE MIGHT SOON BE SIG-
NIFICANT ITEMS i SHOULD TA<E UP DIRECTLY WITH PRES AND IT ADVIS-
ABLE TO SAVr TICKET FOR SUCH AN OCCASION. LETTER WAS DELIVFPED TO 
FONMIN THIS MORNING* BELTON 
LIMITED OFFICIAL USE
\end{verbatim}
\end{mdframed}
\end{small}
\caption{Example of OCR text for the first page of a document.}
\label{fig:ocrtext}
\end{figure}

To simulate the “see a document” query, we use the OCR text from the document that we picked as the basis for the query, and see if the system can guess which box it was from.  Figure~\ref{fig:ocrtext} shows an example of the OCR text from the first page of a document.  We call this the query-by-example scenario.  Note that this results in a rather long query; on average OCR produced 228 words per page.\footnote{For efficiency reasons, in an actual search system we would also want to do some query term selection (e.g.,~\cite{paik2014fixed}).} To limit the complexity of our result tables in this paper, we consistently report results for some number of pages of OCR text that are used both to form the query-by-example and to build the index (e.g., for a box index built from the first page of several documents, our query is built using only the first page of the query-by-example document), although that need not be the case in practice.  

\subsection{Ranking the Boxes}

Whichever type of query we get, we then use bag of words retrieval, ranked with Okapi BM-25 term weights~\cite{DBLP:journals/ftir/RobertsonZ09} (with $k_1=1.2$ and $b=0.75$), with the Porter stemmer, to create a ranked list of the thirty five boxes, hopefully with the correct box at or near the top.\footnote{We used our own BM-25 implementation, included in code distributed with our data.}  As our measure of success, we count how many times our system (which we call ``BoxFinder'') guesses the correct box (i.e., the box that actually contains the document the query was made from).  To easily compute a percentage, we choose 100 query documents and report how many of that 100 the system got right.  We can do this using any number of documents to describe each box, for any number of pages from those documents, and for either way of making a query.  For example, as Table 2 shows, if we make the query from the OCR words on the first page of the PDF file, and we use OCR words from the first page of three (other) PDF files to describe each box, then the system is right 27.9\% of the time.  We call this way of measuring BoxFinder's results Top-1 (since it is the percent of the time that BoxFinder places the correct answer at rank 1).  If it were just guessing randomly, with no real idea which box to look in, it would only be right at Top-1 2.9\% of the time (i.e., once in every 35 tries).  

From this we can conclude BoxFinder is well named – it can find boxes.  Of course, 27.9\% is a long way from perfect, but it need not be perfect to be useful.  Indeed, even when BoxFinder's top result isn't quite right, the right answer is ``close'' more often than chance would predict.  For example, if we look at when BoxFinder's first or second guess is perfect (Top-2), one of those is right 40.4\% of the time when we have three first-page training samples per box.  Moreover, if we ask how often BoxFinder's Top-1 result is within one box numerically (e.g., ranking box 1903, 1904 or 1905 first when it should have found box 1904), that happens 36.8\% of the time with three first-page training samples per box (and we note that this additional benefit from looking ''nearby'' was obtained despite there being gaps in the box numbering in the test collection we have used). 

Of course, guessing randomly is a low baseline.  We can instead index terms generated from the labels on the folders in each box.  To do this, we must decode subject-numeric codes.  The State Department developed a classification guide to help their staff assign codes consistently~\cite{subjectnumeric1963,subjectnumeric}. It is straightforward to replace each code (e.g., POL 12-6) with the corresponding labels (in that case, ''POLITICAL PARTIES: Membership. Leaders.'').\footnote{The 1963 handbook~\cite{subjectnumeric1963} was updated in 1965~\cite{subjectnumeric}. A few codes had different labels in the two; in such cases, we combined terms for that code from both handbooks.}  That's a combination of the label for POL 12 (``POLITICAL PARTIES'') and the label for POL 12-6 (''Membership. Leaders.'') because the State Department classification guide subheadings (in this case, POL 12-6) are meant to be interpreted in the context of the corresponding main heading (in this case, POL 12).  

Subject-Numeric codes sometimes also include abbreviations of the names of countries (e.g., PAR for Paraguay, USSR for the Soviet Union, and US for the United States of America), so we also extract and expand those names to a single standard form (e.g., we do not also expand USSR to Union of Soviet Socialist Republics).  However, we do not extract or expand ``BRAZ'' (Brazil), which appears in every folder label in our collection, since its presence everywhere would result in it having no beneficial effect on the ranking of boxes.  The folder labels also include dates, from which we extract and include the year (e.g., 1964) with the metadata that we index.  We elected not to extract and include the month or day from the date because the distribution of dates that we observe makes it clear that these are start dates for a folder, but that a folder can contain documents from several months.  We also elected not to index the subject-numeric code itself, since we did not expect the queries used in our experiments to contain such codes (although we note that in a practical application, expert searchers might indeed understand and use subject-numeric codes).

We can also optionally include any scope note text.  For example, the scope note for POL 12-6 in the classification guide states ``Includes party elections, purges of party, etc. Subdivide by name of leader if volume warrants.''  Scope notes contain both on-topic terms (e.g., ``party elections'') and off-topic terms (e.g., ``subdivide''), and scope notes can even include negated terms (e.g., the scope note for POL-12 states, in part, ``Exclude: Materials on ... legislative matters, for which SEE: POL 15 -2'').  Experience from the Text Retrieval Conference (TREC) suggests removing negated terms may have little effect on average, since although retaining negated terms is sometimes harmful, they can also sometimes be helpful (because negated terms often have some relation to the topic)~\cite{robertson1999okapi}.\footnote{The TREC experiments compared retention to removal of negated query terms, but because ranking relies on term matching we would expect similar results for retention or removal of content terms from the items being indexed.}  Scope notes can also contain guidance not related specifically to the topic (e.g., ``if volume warrants''), but because such guidance is found in many scope notes, those common terms should have little effect on the way boxes are ranked.  For these reasons, when we include scope notes, we use the full scope note, with no human editing.  Unlike the labels, where we combine the labels for broad topics (e.g., POL 12) and subtopics (POL 12-6), we use only one scope note (in this case, for POL 12-6) because some topic-level scope notes indicate when the topic-level category should be assigned in preference to a subcategory.

\section{Evaluation Measure}

How well the system ranks boxes depends on which documents describe each box, and which documents we pick as queries.  We pick documents to represent each box randomly from within each box, without replacement.  To select query documents, we first randomly select the box the query document will be found in.  We do this 100 times, with replacement.  Then for each of those 100 choices of boxes, we randomly select a query document from that box, being careful not to choose documents that were used to describe that box.  This can choose the same query document twice in a set of 100, but such cases are rare.  Because of these random choices, BoxFinder evaluation scores will vary depending on the choices that we made, so we run the whole experiment 100 times (randomly choosing the documents to represent each box again, and randomly choosing query documents again), averaging those results to get the percentages we report.   

Code and data for the experiments is at \url{https://github.com/oard/BoxFinder}.

\section{Results}

We first look at the case in which boxes are represented using sampled OCR text.  Following that, we look at how the results would differ if folder labels could also be used as a basis for representing the content of a box.

\subsection{Searching Sampled OCR Text}

Table~\ref{tab:short} summarizes the results for title metadata queries.  As we might expect, having a larger number of randomly sampled documents (``samples'') to represent a box yields better results.  Surprisingly perhaps, it’s generally best to use only the first page of each document.  One reason for this might be that some documents are very short---32\% have only a single page---so we only get more pages from those that are longer.

\begin{table*}[t]
\caption{Results using title metadata queries to search OCR from sampled documents.  Top-1: \% in finding box at rank 1.  Top-2: \% finding exact box at rank 1 or 2.}
\centering
\begin{tabular}{c|rr|rr|rr|rr|rr|}
\cline{2-11}
& \multicolumn{2}{|c|}{First Page} & \multicolumn{2}{|c|}{$\le$ 2 Pages} & \multicolumn{2}{|c|}{$\le$ 3 Pages} & \multicolumn{2}{|c|}{$\le$ 4 Pages} & \multicolumn{2}{|c|}{$\le$ 100 Pages}\\ 
\cline{2-11}
Samples & Top-1& Top-2& Top-1& Top-2& Top-1& Top-2& Top-1& Top-2& Top-1& Top-2\\ 
\hline
      1 & 10.7 & 16.9 & 10.5 & 17.1 &  9.5 & 15.5 &  9.3 & 15.5 &  9.2 & 15.4 \\
      2 & 13.0 & 21.1 & 12.2 & 19.4 & 10.8 & 18.1 & 11.6 & 18.6 & 10.9 & 18.0 \\
      3 & 14.6 & 22.1 & 14.1 & 21.5 & 12.9 & 20.5 & 13.0 & 20.2 & 11.7 & 19.2 \\
      4 & 15.6 & 23.6 & 14.9 & 23.2 & 13.8 & 21.0 & 13.6 & 21.4 & 12.9 & 20.3 \\
      6 & 16.9 & 25.4 & 15.8 & 24.3 & 13.7 & 22.4 & 13.8 & 21.9 & 12.2 & 20.6 \\
      8 & 16.6 & 25.0 & 16.0 & 24.5 & 14.8 & 23.1 & 14.4 & 22.6 & 12.9 & 21.3 \\
     10 & 18.1 & 27.1 & 15.6 & 24.5 & 15.0 & 23.8 & 15.0 & 23.5 & 13.3 & 22.2 \\
\hline
\end{tabular}
\label{tab:short}
\end{table*}

\begin{table*}[t]
\caption{Results using query-by-example to search OCR text from sampled documents, using the same page limit for queries and for sampled documents.}
\centering
\begin{tabular}{c|rr|rr|rr|rr|rr|}
\cline{2-11}
& \multicolumn{2}{|c|}{First Page} & \multicolumn{2}{|c|}{$\le$ 2 Pages} & \multicolumn{2}{|c|}{$\le$ 3 Pages} & \multicolumn{2}{|c|}{$\le$ 4 Pages} & \multicolumn{2}{|c|}{All Pages}\\ 
\cline{2-11}
Samples & Top-1& Top-2& Top-1& Top-2& Top-1& Top-2& Top-1& Top-2& Top-1& Top-2\\ 
\hline
      1 & 16.5 & 24.5 & 15.2 & 22.7 & 15.2 & 22.1 & 14.3 & 21.5 & 10.4 & 16.9 \\
      2 & 23.4 & 34.4 & 21.1 & 29.3 & 18.4 & 27.7 & 18.6 & 26.9 & 13.0 & 19.4 \\
      3 & 27.9 & 40.4 & 25.6 & 34.9 & 21.8 & 32,0 & 22.2 & 30.5 & 17.2 & 25.5 \\
      4 & 31.5 & 43.7 & 25.0 & 36.2 & 24.1 & 35,8 & 24.6 & 34.8 & 20.0 & 28.1 \\
      6 & 34.1 & 47.4 & 29.7 & 42.3 & 28.6 & 40.0 & 27,2 & 39.2 & 12.5 & 32.4 \\
      8 & 35.0 & 49.0 & 33.6 & 46.7 & 31.3 & 44.1 & 29.4 & 40.5 & 26.8 & 38.6 \\
     10 & 39.2 & 53.5 & 34.7 & 47.2 & 33.0 & 46.0 & 32.1 & 43.9 & 27.4 & 38.8 \\
\hline
\end{tabular}
\label{tab:long}
\end{table*}

\begin{figure}[t!]
\centering
  \begin{center}
      \includegraphics[trim={1.0in 1.2in 1.0in 1.0in},clip,width=0.85\textwidth]{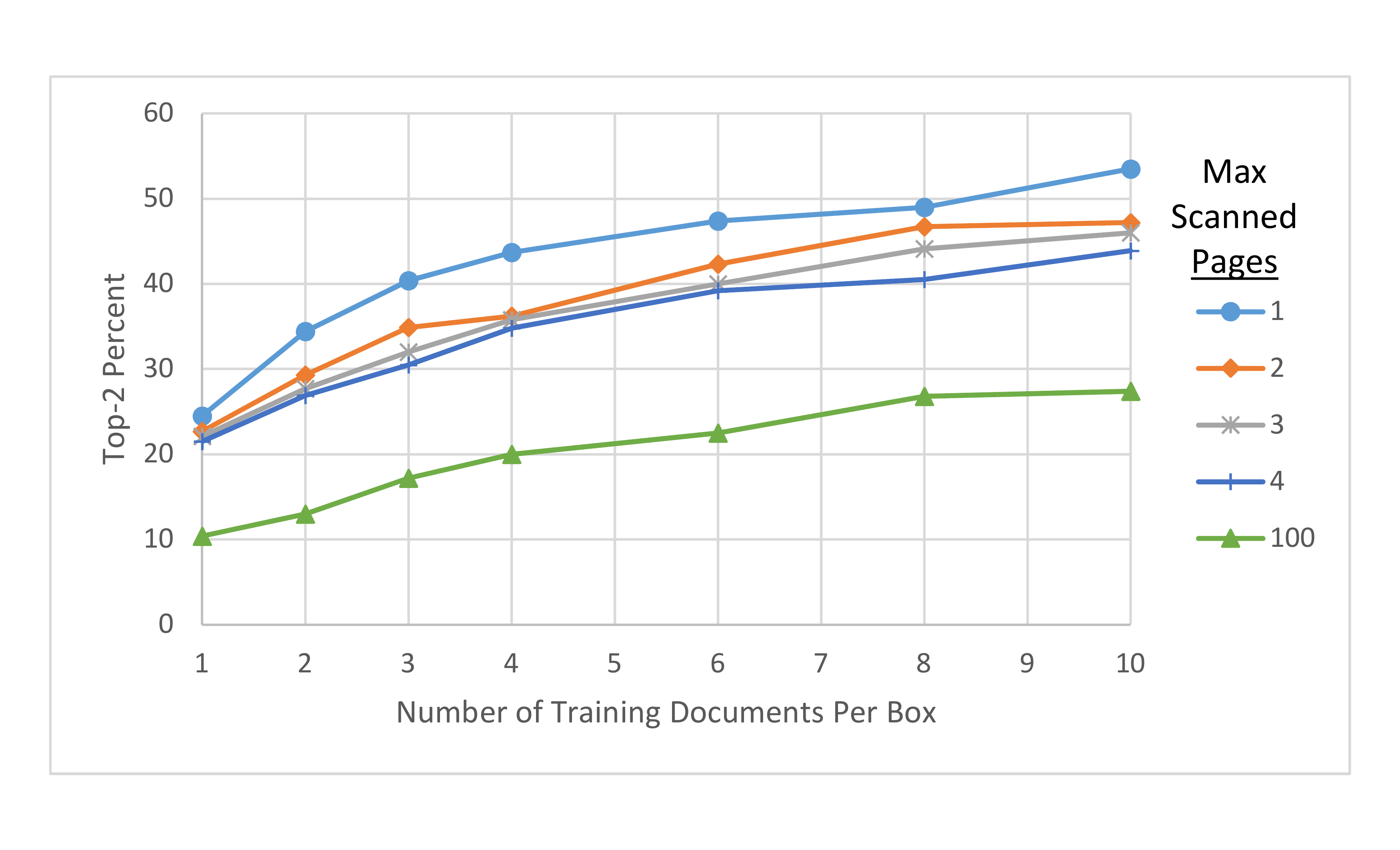}
  \end{center}
  \caption{Learning rate for correct box in Top-2 as scanned documents per box grows, query-by-example condition.  Scanning just the first page is best.}
  \label{fig:learning}
\end{figure}

Table~\ref{tab:long} summarizes results for queries built using OCR text from the query document, and Figure~\ref{fig:learning} illustrates those results for the Top-2 condition. As can easily be seen, BoxFinder does better with these longer queries.  One reason for this is that with short queries (like the ones from title metadata) BoxFinder sometimes finds no matching terms at all, but that doesn't happen very often with longer queries that are based on full-text OCR.  Of course, these longer queries have all kinds of strange things in them (from letterhead, message addresses, OCR errors, handwriting that gets misrecognized, etc.), and the document representations suffer from the same problem.  Nonetheless, there clearly is a lot of signal here in the midst of all that noise, since BoxFinder is doing better with these longer queries than with the title metadata queries, and it is doing way better than random guessing with either of them.  

\begin{table*}[b!]
\caption{Results for searching terms generated from folder labels.}
\centering
\begin{tabular}{l|c|rr|rr|}
\cline{3-6}
\multicolumn{2}{c|}{} & \multicolumn{2}{c|}{No Scope Notes} & \multicolumn{2}{|c|}{With Scope Notes} \\ 
\cline{1-6}
\multicolumn{1}{c|}{Queries} & Repetitions & Top-1& Top-2& Top-1& Top-2 \\ 
\hline
Short: Title metadata   & 4,200 & 12.4 & 17.1 & 12.4 & 18.6 \\
Long: First-page OCR &   100 &  5.0 &  9.5 &  5.3 &  9.0 \\
\hline
\end{tabular}
\label{tab:metadata}
\end{table*}

\begin{table}[t]
\caption{Reciprocal rank fusion results, title metadata queries, merging results from searching terms generated from folder labels (no scope notes) and from searching OCR.}
\centering
\begin{tabular}{c|rr|rr|rr|rr|rr|}
\cline{2-11}
& \multicolumn{2}{|c|}{First Page} & \multicolumn{2}{|c|}{$\le$ 2 Pages} & \multicolumn{2}{|c|}{$\le$ 3 Pages} & \multicolumn{2}{|c|}{$\le$ 4 Pages} & \multicolumn{2}{|c|}{$\le$ 100 Pages}\\ 
\cline{2-11}
Samples & Top-1& Top-2& Top-1& Top-2& Top-1& Top-2& Top-1& Top-2& Top-1& Top-2\\ 
\hline
      1 & 14.0 & 21.5 & 14.0 & 21.8 & 13.3 & 21.0 & 13.1 & 20.5 & 13.3 & 20.3 \\
      2 & 15.5 & 23.4 & 14.8 & 22.7 & 13.9 & 21.8 & 14.8 & 22.6 & 14.5 & 23.0 \\
      3 & 15.8 & 23.5 & 15.5 & 23.7 & 15.1 & 23.4 & 15.1 & 23.0 & 14.5 & 23.0 \\
      4 & 16.6 & 25.2 & 16.8 & 24.5 & 16.2 & 24.1 & 16.5 & 24.0 & 15.0 & 23.1 \\
      6 & 17.6 & 25.7 & 17.2 & 25.1 & 15.8 & 23.9 & 16.1 & 24.1 & 15.1 & 23.5 \\
      8 & 17.7 & 25.9 & 17.3 & 25.3 & 17.7 & 25.6 & 16.6 & 23.8 & 15.4 & 23.5 \\
     10 & 19.1 & 27.5 & 17.4 & 25.7 & 17.1 & 25.1 & 16.7 & 25.1 & 16.1 & 24.6 \\
\hline
\end{tabular}
\label{tab:fusion}
\end{table}

\subsection{Searching with Folder Labels}

Table~\ref{tab:metadata} shows that when short (title metadata) queries are used to search documents that are represented \underline{only} using terms generated from the folder labels, the results are comparable to those reported in Table~\ref{tab:short} for using the same queries to search a single page of OCR-generated text.  The Top-1 results for searching terms generated from folder labels using short queries was 12.4\%, and at Top-2 that same condition had the correct answer in rank 1 or rank 2 17.1\% of the time.  Only about 70\% of the short queries have at least one query term that matches any term resulting from expanding the subject-numeric codes found in the folder labels (without also indexing the scope notes), but short (title metadata) queries still did far better than the longer OCR-based queries when searching document representations that are based solely on terms generated from folder labels.  

Interestingly, the pattern in Tables~\ref{tab:short} and~\ref{tab:long}, where query-by-example was markedly better than using the title metadata as a query, is reversed when ranking based on folder labels.  Essentially the broader pattern we see is that a matched condition (using OCR to search OCR, or using metadata to search metadata) is consistently outperforming an unmatched condition.  This might be explained by systematic errors in the OCR or by systematic differences in the way language is used in the documents and in Brown University's title metadata.  We see a benefit at rank 2 from the inclusion of scope notes when generating terms from folder labels to represent each box, but no benefit at rank 1.\footnote{To measure the benefit of scope notes with short titles more accurately, we average over 4,200 repetitions for our short-query experiments in Table~\ref{tab:metadata}.}

Of course, we need not index terms generated from the folder labels in isolation---we can also use folder labels together with OCR text from sampled documents.  We could do this in one of two ways, either concatenating the two representations, or performing two searches (one with each representation) then performing result fusion to create a single ranked list of boxes.  We expect that second approach, implemented as reciprocal rank fusion~\cite{cormack2009reciprocal}, to work better in this case because of mismatched document lengths, so that's the one we tried.\footnote{We set Cormack's discount rate parameter to 60, as Cormack recommends~\cite{cormack2009reciprocal}.} 

Table~\ref{tab:fusion} shows reciprocal rank fusion results when (short) title metadata is used as the query.  As with Tables~\ref{tab:short} and~\ref{tab:long}, these are averages over 100 repetitions. With these short queries, sometimes no terms match at all, resulting in no ranking of the boxes by one of the systems.  In such cases, we retain the other ranking unchanged.\footnote{When neither approach has a term match, we generate an empty list.}  As can be seen by comparing Tables~\ref{tab:metadata} and~\ref{tab:fusion}, this rank fusion results in an improvement over what we achieved using terms generated from folder labels alone.  This improvement is both substantial (for example, compare 15.8 at Top-1 for Reciprocal Rank Fusion with the first page from each of 3 samples in Table~\ref{tab:fusion} to to 12.4 for folder metadata alone in Table~\ref{tab:metadata}, a 27\% relative improvement; the relative improvement at Top-2 is 38\%) and statistically significant (the standard deviation over 42 100-sample averages when searching terms generated from folder labels is 0.33 at both Top-1 and Top-2). 

\section{Conclusion and Future Work}

We close by observing that we have shown that the homophily between digitized and undigitized content that we expected to find in an archival collection can indeed provide a useful signal that can help to improve search for content that has not yet been digitized.  There are several ways in which we might push this work further.  One thing to try would be to be selective about which parts of a document image to index.  For that, we could pay attention not just to the OCR, but to cues from the layout of the words on the page.  For example, we might pay particular attention to who sent or received a document, or to the date of the document.  We could also use layout analysis to determine what type of document we are looking at (e.g., telegram, letter, memo, or form), and then apply type-specific information extraction, and possibly even type-specific ranking.  Speaking of ranking, there’s no reason why we need to glom the OCR text from different documents together to make a single representation for each box.  Instead, we could make multiple representations, one per document, and then let those representations vote on which box should be chosen.  That approach has, for example, been shown to work well for blog search~\cite{balog2006formal}.  

There is nothing in BoxFinder specific to boxes except the way we tested it; the same ideas could work for folders, series, collections, and entire repositories.
Of course, some of the tuning decisions (e.g., how many digitized documents are needed to represent a folder?) will likely differ when applying these ideas at different scales.  But tuning would not be hard if we had a collection to tune on. So one key to making BoxFinder into FolderFinder (or SeriesFinder, or …) is to assemble appropriate collections on which we can train and test.  For our experiments in this paper, we assembled a single collection and then used it to see how well our approach of representing a box using OCR from randomly selected documents would do.  But when tuning a large number of system details, we’ll want training, devtest and evaluation partitions, so we'll need larger collections.  Fortunately, the complete NARA Department of State Subject-Numeric files are indeed much larger than the part of that collection that we have used so far, so there is at least one good source for such a collection.  But if we want to know how well these ideas work in general, we’re going to need several collections, from a variety of sources.  
So assembling several such collections is a natural next step.

Finally, we have looked only at what can be done using systematic sampling of densely digitized documents, together with quite terse folder-level metadata, using just one test collection.  Future work should explore other cases, where the degree of homophily within a box (or other unit) may vary, the available metadata that describes the content is richer (or less rich), and digitization is unevenly distributed across the collection, as is often the case in practice.  

\section*{Acknowledgements}

The author greatly appreciates the comments on this work from from Katrina Fenlon, Emi Ishita, Diana Marsh, Toshiyuki Shimizu, Tokinori Suzuki, Yoichi Tomiura, Victoria Van Hyning, and the reviewers.

\bibliographystyle{splncs04}
\bibliography{tpdl}

\end{document}